\documentclass[12pt]{article}

\usepackage[totalheight = 21cm, totalwidth = 17cm]{geometry}
\usepackage{amssymb,amsmath,amsfonts,amsbsy,graphicx}
\usepackage{color}

\newcommand{\dis}[1]{\begin{equation}\begin{split}#1\end{split}\end{equation}}

\begin{document}

\begin{titlepage}


\rightline{\footnotesize{MAD-TH-17-10}} \vspace{-0.2cm}

\begin{center}

\vskip 1.0 cm

{\Large \bf Electromagnetic Duality and the Electric Memory Effect

}

\vskip 1.0cm

{\large
Yuta Hamada$^{a, b}$,
 Min-Seok Seo$^{c}$,
 Gary Shiu$^{a}$ 
}

\vskip 0.5cm

{\it
$^{a}$ Department of Physics, University of Wisconsin-Madison, Madison, Wisconsin 53706, USA
\\
$^{b}$ KEK Theory Center, IPNS, KEK, Tsukuba, Ibaraki 305-0801, Japan
\\
$^{c}$Department of Physics, Chungnam National University, Daejeon 34134, Korea
}

\vskip 1.2cm

\end{center}

\begin{abstract}

 We study large gauge transformations for soft photons in quantum electrodynamics which, together with the helicity operator, form an $ISO(2)$ algebra.
 We show that the two non-compact generators of the $ISO(2)$ algebra 
 correspond respectively to the residual gauge symmetry and its electromagnetic dual gauge symmetry that emerge at null infinity.
 The former is helicity universal  (electric in nature) while the latter is helicity distinguishing (magnetic in nature). 
Thus, the conventional large gauge transformation is electric in nature, and is naturally associated with a scalar potential. 
 We suggest that the electric Aharonov-Bohm effect is a direct measure for the electromagnetic memory arising from large gauge transformations.

\end{abstract}

\end{titlepage}

\newpage

\section{Introduction}
\setcounter{equation}{0}

The physics of soft photons and gravitons has drawn much attention  since the 1960s as the soft limit can reveal many aspects of gauge theory and gravity that are not easily accessible by massive particles.
 One famous folklore is that the infrared (IR) divergences of soft photon and graviton are governed by gauge symmetries \cite{Weinberg:1964ew} and  thus factorized from other hard processes \cite{Weinberg:1965nx}.
 Recently,  such `soft theorems' were given a new interpretation 
 in terms of the asymptotic symmetries (for a review and summary of earlier works, see \cite{Strominger:2017zoo}).
 It is motivated by the observation that while gauge redundancy is necessarily eliminated through gauge fixing, a part of the gauge transformation 
  emerges as an asymptotic symmetry to an observer far away from the source.
 These `large gauge transformations' (LGTs) are well studied in the asymptotic flat spacetime, where appropriate fall-off boundary conditions permit gauge transformation parameters depending only on the angular coordinates to become approximate symmetry around null infinity.
The  LGTs connect different solutions of the classical equations of motion under the given boundary conditions, or equivalently, degenerate vacuum configurations corresponding to different coherent 
excitations
of soft photons/gravitons.
 As a result, the LGT generators create or annihilate soft photons/gravitons, and the associated 
 soft theorem can be interpreted as a Ward identity of the LGT.
 Indeed, the change of the vacuum configuration after soft photon/graviton emission gives rise to 
 physically observable effects known as the electromagnetic/gravitational memory.
 The common structure shared by the memory effect and the soft theorem, together with the fact that the change of the vacuum configuration from the memory effect is described by a LGT, led one to conjecture  a  `triangular relation' 
 that governs the IR  properties of gauge theory and gravity \cite{Strominger:2017zoo}. 
 The  `triangular relation' holds even in the scalar theory~\cite{Campiglia:2017dpg,Hamada:2017atr,Campiglia:2017xkp}.
 
 The triangular relation for gravity has been extensively studied in the context of asymptotic flat background, where the LGT is  known as the Bondi-van der Burg-Metzner-Sachs (BMS) transformation \cite{Bondi:1962px, Sachs:1962wk, Sachs:1962zza}.
 For more generic backgrounds,  different spacetime structures with e.g. the absence of light-like null infinity or the existence of a horizon makes closure of the triangular relation more challenging. 
Yet, some of these IR relations seem to hold even in cosmological backgrounds \cite{Bieri:2015jwa, Kehagias:2016zry, Mirbabayi:2016xvc, Ferreira:2016hee, Hamada:2017gdg, Tanaka:2017nff} where the spacetime is not asymptotically flat. 
 On the other hand, the conceptual and technical difficulties arising from the close connection between the background  spacetime and the LGTs do not exist in Abelian gauge theory like quantum electrodynamics (QED) in flat spacetime \cite{He:2014cra, Kapec:2015ena}.
 In QED, the photons at null infinity distinguish the two different helicity states.
 Since the helicity structure of the photon originates from the representation of the Lorentz group for massless particles, 
 it is natural to associate the LGT with the structure of the Lorentz group.
 More concretely, the helicity states correspond to the irreducible representation of the little group, a subgroup of the 
 Lorentz group that does not change the momentum of particle.
 For a massless particle, the little group is given by $ISO(2)$, which contains not just a helicity operator, but also two non-compact generators.
 When acting on a photon, these two non-compact generators create a longitudinal mode in a helicity-dependent and a helicity-independent  way, respectively, which have been interpreted as gauge transformations.
 Interestingly, the LGT takes the form of a helicity-independent gauge transformation at null infinity.
 Moreover, we find that in order to close the algebra describing the soft photon around null infinity, we need another infinite-dimensional generator corresponding to the large distance limit of a helicity-dependent gauge transformation.
 Then three symmetry charges of the soft photon: its helicity, LGT charge and the newly introduced generator form the $ISO(2)$ algebra \cite{Hamada:2017uot}. 
 
 In this article, we study in detail such helicity related issues of the LGT and the non-compact generator of $ISO(2)$.
 In Sec. \ref{sec:interpret}, we provide an interpretation of the new non-compact generator in light of electromagnetic duality.
 In the absence of massive charged particles, only soft photons propagate at null infinity, whose equations of motion are given by the source-free Maxwell equations locally.
 Hence, we expect that electromagnetic duality introduces a  `dual gauge field' $\widetilde{A}$ defined by $\widetilde{F}=d \widetilde{A}$, where $\widetilde{F}$ is the dual electromagnetic field strength whose components are given by $\widetilde{F}_{\mu\nu}=\frac12\epsilon_{\mu\nu\rho\sigma}F^{\rho\sigma}$.
 We claim that the dual gauge transformation $\widetilde{A} \to \widetilde{A}+d\Lambda$ is a helicity-distinguishing gauge transformation, whose long distance limit is given by the action of the other non-compact $ISO(2)$ generator on the soft photons.
 In Sec. \ref{Sec:AB}, hinted by the helicity-independent nature of  the LGT charge, a property shared by the electric field, we propose the `electric Aharonov-Bohm effect' as a way to observe the memory effect in QED.
It was previously suggested that the electromagnetic memory effect can be observed through the well-known magnetic
Aharonov-Bohm effect \cite{Susskind:2015hpa} by measuring the vector field in different gauges.
 It is interesting to point out that the original Aharanov-Bohm paper \cite{Aharonov:1959fk} also mentioned  the electric Aharonov-Bohm effect, as a way to measure the electric potential difference. 
 We show here that the LGT generator with a delta functional gauge transformation parameter can be interpreted as an electric potential, so the electric Aharonov-Bohm effect may be a more direct way to measure the electromagnetic memory.

\section{Completion of the LGT Algebra and its Interpretation}
\label{sec:interpret}
\setcounter{equation}{0}                                                                                                                                                                                                                                                                                                                                                                                                                                                                                                                                                                                                                                      

\subsection{LGT around null infinity}

In this section, we review how the LGT emerges around null infinity and the vector field transforms under the LGT.
 Since gauge symmetry is a redundancy of physical degrees of freedom, the quantization of gauge field requires gauge fixing, i.e., choosing only one `orbit' of the physical degrees of freedom through the condition $G(A_\mu)=0$.
 All the redundancies are completely eliminated provided that Faddeev-Popov determinant does not vanish,
\dis{{\rm Det}\Big[\frac{\delta G}{\delta A_\mu}\partial_\mu\Big]\ne 0.}
In the case of Abelian gauge theory such as QED, we need one gauge fixing condition for each spacetime point, so in principle, the `residual gauge transformation', the gauge transformation surviving the gauge fixing does not exist.
 Instead, even though not exact, an approximate residual gauge symmetry can emerge at some specific region \cite{Avery:2015rga}.
 More specifically, a residual gauge symmetry that emerges at null infinity is called a large gauge transformation (LGT).
 Therefore, in the gauge fixed Lagrangian,
  \dis{{\cal L}=-\frac14 F_{\mu\nu}F^{\mu\nu}+\frac{1}{2\xi}B^2+ B\partial_\mu A^\mu-\overline{c}\partial^2 c,}
  where $B$ is the auxiliary field used to impose the gauge fixing condition and $c, \overline{c}$ are ghosts,
  while gauge transformation appears as a BRST transformation,
  \dis{\delta A_\mu=\epsilon\partial_\mu c,\quad\delta c=0,\quad \delta\overline{c}=\epsilon B,\quad\delta B=0,\label{Eq:BRST}}
 the  LGT  is just an accidental symmetry which holds around null infinity, under which
\dis{\delta A_\mu =\partial_\mu \varepsilon,\quad  \delta c=0,\quad \delta\overline{c}=0,\quad\delta B=0,\label{Eq:LGT}}
making $\delta {\cal L}(r\to \infty)={\cal O}(r^{-n})$ for some positive $n$.

 To see the situation explicitly, we consider the  `retarded time' coordinate $(u, r, z, \bar{z})$ where the retarded time $u=t-r$ and the complex  parametrizations for the angular variables $z=\tan(\theta/2)e^{i\phi}$ and $\bar{z}$ are used, with the flat spacetime metric given by 
\dis{&ds^2=-du^2-2 du dr+2r^2\gamma_{z\bar{z}}d\bar{z}dz,\quad\quad \gamma_{z\bar{z}}=\frac{2}{(1+\bar{z}z)^2}.}
Taking  the Lorenz gauge, $G=\partial_\mu A^\mu=0$, for example, the LGT forming the residual gauge symmetry is parametrized by the solution of the  `zero mode equation' $\varepsilon(x)$,
\dis{\frac{\delta G(A)}{\delta A_\mu}\partial_\mu \varepsilon(x)=\square \varepsilon(x)=0.}
Under the boundary condition $\lim_{r\to\infty}\varepsilon(x)=0$, we only have a trivial solution $\varepsilon=0$.
On the other hand, when we loosen the boundary condition with an appropriate falloff behavior, say,
\dis{\lim_{r\to\infty}A_u={\cal O}(r^{-1}),\quad
\lim_{r\to\infty}A_r={\cal O}(r^{-2}),\quad
\lim_{r\to\infty}A_{z/\bar{z}}={\cal O}(1),\label{Eq:boundary}}
the parameter $\varepsilon$ may solve the zero mode equation allowing ${\cal O}(r^{-n})$ ($n>0$) corrections which are not so important around null infinity.
Hence, expanding $\varepsilon$ in terms of $r$
\footnote{ See also \cite{Campiglia:2015qka, Strominger:2017zoo} for the solution of the zero mode equation.},
\dis{\varepsilon(x)=\varepsilon^{(0)}(u, z, \bar{z})+\frac{1}{r}\varepsilon^{(1)}(u, z, \bar{z})+\frac{1}{r^2}\varepsilon^{(2)}(u, z, \bar{z})+\cdots , }
and inserting this into the zero mode equation,
\dis{0&=\square \varepsilon=\frac{1}{\sqrt{-g}}\partial_\mu(\sqrt{-g}g^{\mu\nu}\partial_\nu\varepsilon)=-\partial_u \partial_r \varepsilon+\frac{1}{r^2}\partial_r[r^2(\partial_r \varepsilon-\partial_u\varepsilon)]+\frac{2}{r^2\gamma_{z\bar{z}}}\partial_z\partial_{\bar z}\varepsilon
\\
&=-\frac{2}{r}\partial_u \varepsilon^{(0)}+\frac{1}{r^2}[\frac{2}{\gamma_{z{\bar z}}}\partial_z \partial_{\bar z}\varepsilon^{(0)}]+\cdots,
}
so far as ${\cal O}(r^{-2})$ terms are neglected, $\partial_u \varepsilon^{(0)}=0$, or $\varepsilon^{(0)}=\varepsilon^{(0)}(z, \bar{z})$ is a good approximation to a residual gauge symmetry, or large gauge symmetry as it approximately solves the zero-mode equation.

 Since the LGT parameter $\varepsilon$ depends only on the angular variables $(z, \bar{z})$, only the angular components of the gauge field transform non-trivially under the LGT : $A_{z/\bar{z}} \to A_{z/\bar{z}}+\partial_{z/\bar{z}}\varepsilon(z, \bar{z})$.
 Indeed, from the behavior of a plane wave at $r\to\infty$ \cite{Strominger:2017zoo,Gabai:2016kuf},
   \dis{\lim_{r\to\infty}e^{-i k\cdot x}=\frac{2\pi i}{\omega r}\Big[e^{i(\omega+i\epsilon)(u+2 r)}\delta^2(\hat{x}+\hat{k})-
   e^{i(\omega+i\epsilon)u}\delta^2(\hat{x}-\hat{k})\Big],\label{Eq:planewave}}
 we find that the photon momentum direction becomes parallel to the direction of the photon's classical trajectory given by 
   \dis{\hat{x}=\frac{1}{1+\bar{z} z}(z+\bar{z}, -i(z-\bar{z}), 1-\bar{z}z).}
  Moreover, the  angular components of the gauge field   $A_z=\partial_z x^\mu A_\mu$ and $A_{\bar z}$ are written as
  \dis{&A_z=-i\sqrt{\gamma_{z\bar{z}}}\int^\infty_0\frac{d\omega}{8\pi^2}(a_+(\omega\hat{x})e^{-i \omega u}-a_-^\dagger(\omega\hat{x})e^{i\omega u}),
\\
&A_{\bar z}=-i\sqrt{\gamma_{z\bar{z}}}\int^\infty_0\frac{d\omega}{8\pi^2}(a_-(\omega\hat{x})e^{-i \omega u}-a_+^\dagger(\omega\hat{x})e^{i\omega u}),\label{Eq:gaugemode}} 
 showing a similar structure as a Weyl fermion.
 That means that in the Weyl representation of a massless fermion, the left-handed Weyl field describes the negative helicity particle and the positive helicity antiparticle whereas the right-handed Weyl field pairs the positive helicity particle with the negative helicity antiparticle.
 Of course, the vector field is real, so the same structure appears that $A_z$ contains the positive helicity photon annihilation operator and the negative helicity photon creation operator while $A_{\bar z}$ contains the negative helicity photon annihilation operator and the positive helicity photon creation operator.
 Hence, the gauge field at null infinity shows its helicity structure explicitly through its complexified angular components. 
 
 Now, as shown in eq. (\ref{Eq:LGT}), LGT corresponds to the replacement of operators $A_{z/\bar{z}}$ with helicity $\pm 1$ by the scalar functions $\partial_{z/\bar{z}}\varepsilon$, contrary to the original gauge transformation which replaces the unphysical polarization states by the ghost (eq. (\ref{Eq:BRST})).
That means, the action of the LGT generator $Q_\varepsilon$ satisfying $i[Q_\varepsilon, A_{z/\bar{z}}]=\partial_{z/\bar{z}} \varepsilon$ changes the helicity of the photon states.
This can be checked explicitly from the expression of the LGT generators, which was found in \cite{He:2014cra, Kapec:2015ena, Gabai:2016kuf}.
 Applying Noether's theorem to the gauge symmetry ($A_\mu \to A_\mu+\partial_\mu \Lambda$), we obtain
\dis{ J_N^\mu=\frac{\partial{\cal L}}{\partial(\partial_\mu A_\nu)} \delta A_\nu + J^\mu \Lambda=-F^{\mu\nu}\partial_\nu \Lambda + J^\mu \Lambda=-(F^{\mu\nu}\Lambda)_{;\nu}+(F^{\mu\nu}_{~;\nu}+J^\mu)\Lambda,}
where the second term vanishes by the equations of motion.
Taking $\Lambda$ to be global, we obtain the conventional electric charge after dropping $\Lambda$.
In the case of the LGT generated by $\varepsilon(z, \bar{z})$, we also have the LGT charges coming from an infinite number of Noether currents, $J_{N,\varepsilon}^\mu=-(F^{\mu\nu}\varepsilon)_{;\nu}$,
\dis{Q_\varepsilon=-\int_S d^2z r^2\gamma_{z\bar{z}} \varepsilon(z, \bar{z})F_{ru}.\label{Eq:LGTcharge}}
Under the boundary conditions given by eq. (\ref{Eq:boundary}), the dominant  terms of the $r$ component of the Maxwell equations around null infinity is written as
\dis{-\partial_u F_{ru}+\frac{1}{r^2\gamma_{z\bar{z}}}[\partial_z(F_{r\bar{z}}-F_{u\bar{z}})+\partial_{\bar z}(F_{rz}-F_{uz})]
\simeq -\partial_u F_{ru}-\frac{1}{r^2\gamma_{z\bar{z}}}\partial_u(\partial_zA_{\bar z}+\partial_{\bar z}A_z)=-J^r. \label{Eq:Maxwell}}
Note that even though eq. (\ref{Eq:Maxwell}) is a part of $\vec{\nabla}\times\vec{B}=\partial \vec{E}/\partial t +\vec{J}$, all the dominant terms come from the electric field.
In particular, the angular components $A_{z/\bar{z}}$ we are interested originated from $F_{u z/\bar{z}}$, which appears due to the choice of the retarded time coordinate.
Integrating eq. (\ref{Eq:Maxwell}) over the spherical surface of null infinity ($r\to \infty$), and again over retarded time, we obtain the change of LGT charge,
\dis{Q_\varepsilon(u=+\infty)&-Q_\varepsilon(u=-\infty)
=-\int^{+\infty}_{-\infty} du\int d^2 z r^2 \gamma_{z\bar{z}}\varepsilon \partial_u F_{ru}
\\
&=-\int^{+\infty}_{-\infty} du\int d^2 z r^2 \gamma_{z\bar{z}}\varepsilon   J^r
+\int^{+\infty}_{-\infty} du\int d^2 z \varepsilon  \partial_u (\partial_zA_{\bar z}+\partial_{\bar z}A_z).\label{Eq:chargecons}}
Hence, the conservation law does not just equate the change of charge with flux (the first term of RHS), but should be supplied by an additional contribution, the second term of the RHS. 
Using eq. (\ref{Eq:gaugemode}), we find that the second term represents the sum of the soft photons on null infinity over $-\infty < u<+\infty$, in the form of
\dis{\Delta Q_\varepsilon^{(s)}=\int d^2z \lim_{\omega\to 0}\frac{\omega \sqrt{\gamma_{z\bar{z}}}}{8\pi}\Big[&(\partial_z\varepsilon(z,\bar{z})a_+ (\omega\hat{x}) +\partial_{\bar z}\varepsilon(z, \bar{z})a_+^\dagger(\omega\hat{x}))
\\
&+(\partial_{\bar z}\varepsilon(z,\bar{z})a_-(\omega\hat{x})+\partial_z\varepsilon(z,\bar{z})a_-^\dagger(\omega\hat{x}))\Big].
\label{Eq:LGTharge}}
For this reason, we call the second term the ``soft part" of the LGT charge, in contrast with the ``hard part" given by the first term.
 In the absence of massless charged particle, only the soft part is considered at null infinity. 
  Since the soft part $\Delta Q_\varepsilon^{(s)}$ is linear in the creation/annihilation operators of soft photon for each helicity and momentum direction, its action on a soft photon state is a linear superposition of states with one more and one less left- and right-handed soft photon \cite{Hamada:2017uot}.
  It is clear that the helicity of these states are different from the original one.

\subsection{Completion of the LGT algebra: $ISO(2)$ structure}
\label{Sec:algebra}

 The fact that the two symmetry generators of the soft photon at null infinity, namely, the helicity operator $J$ and the soft part of LGT charge $\Delta Q_\varepsilon^{(s)}$ do not commute with each other indicates that more charge(s) need to be introduced to close the algebra.
 As shown in  \cite{Hamada:2017uot}, the minimal closed algebra is $ISO(2)$ which requires one more generator.
 To see this, we notice that $J$ and $\Delta Q_\varepsilon^{(s)}$ has the structure of
 \dis{&J=\int d^2z \big[ a_{+, z}^\dagger a_{+, z}-a_{-, z}^\dagger a_{-, z} \big],
 \\
 &\Delta Q_\varepsilon^{(s)}=\int d^2z \sum_{ \lambda=\pm} g(z, \bar{z}) \big[ (a_{+, z}+a^\dagger_{+, z})+ (a_{-, z}+a^\dagger_{-, z})\big], \label{Eq:JandQ}}
 where the commutation relations are given by
 \dis{[a_{\lambda,z}, a^\dagger_{\lambda', z'}]=\delta_{\lambda, \lambda'}\delta^2(z- z'),
 \quad
[a_{\lambda,z}, a_{\lambda', z'}]=0=[a^\dagger_{\lambda,z}, a^\dagger_{\lambda', z'}].}
 Moreover, in order to define a common, real weight function $g(z, \bar{z})=(\omega\sqrt{\gamma_{z\bar{z}}}/8\pi)|\partial_z \varepsilon|$, we absorb the phase of $\partial_z \varepsilon$ into $a_{\lambda, z}$.
 Obviously, for a fixed angle $z$, $\Delta Q_\varepsilon^{(s)}$ has the structure of the position operator for the two-dimensional simple harmonic oscillator, corresponding to the two distinct helicities.
 Hence, if we introduce another operator $\Delta P_\varepsilon^{(s)}$ given by
 \dis{\Delta P_\varepsilon^{(s)}=\int d^2 z \sum_{ \lambda=\pm} g(z, \bar{z})\big[ i(a_{+, z}-a^\dagger_{+, z})-i(a_{-, z}-a^\dagger_{-, z})\big],\label{Eq:P}}
 three operators $J$, $\Delta Q_\varepsilon^{(s)}$, and $\Delta P_\varepsilon^{(s)}$ satisfy the $ISO(2)$ algebra,
 \dis{[\Delta Q_\varepsilon^{(s)}, \Delta P_\varepsilon^{(s)}]=0,\quad
 [J, \Delta Q_\varepsilon^{(s)}]=i \Delta P_\varepsilon^{(s)},\quad
 [J, \Delta P_\varepsilon^{(s)}]=-i \Delta Q_\varepsilon^{(s)}.}
Interestingly, as pointed out in \cite{Hamada:2017uot},  the $ISO(2)$ algebra also appears in the representation of the Lorentz group.
 More precisely, instead of the non-compact Lorentz group, we take its subgroup that does not alter the momentum of the state, such that a `particle' is defined in terms of a discrete, finite dimensional representation of such `little group' \cite{Wigner:1939cj}.
 Whereas the little group for a massive particle is given by $SO(3)$, generated by the spin operators, the little group of the massless particle is given by $ISO(2)$.
 In the $ISO(2)$ algebra, besides the helicity operator $J$, we have two non-compact generators $\Pi_1$ and $\Pi_2$ satisfying
  \dis{[\Pi_1, \Pi_2]=0,\quad [J, \Pi_1]=i\Pi_2, \quad [J, \Pi_2]=-i\Pi_1.}
 Since $\Pi_{1,2}$ give continuous and infinite eigenvalues, 
 the common practice is to
  fix them to specific values, and take the helicity as a quantum number labelling the particle state.
 However, as pointed out in \cite{Weinberg:1964ew}, the action of $\Pi_{1,2}$ on the vector field has the same form as the gauge transformation.
 This can be explicitly checked by considering the photon moving in the $z$-direction. 
 In this case, the helicity operator is just $J_3$, and the other two non-compact generators are given by $\Pi_1=J_2+K_1$ and $\Pi_2=-J_1+K_2$, which induce the non-compact little group transformation
 \dis{W(\alpha, \beta)={\rm exp}[i(\alpha \Pi_1+\beta \Pi_2)]=\left(
\begin{array}{cccc}
1+\frac{\alpha^2+\beta^2}{2} & \alpha & \beta & -\frac{\alpha^2+\beta^2}{2}   \\
\alpha & 1 & 0 &-\alpha \\
\beta &0 & 1 &-\beta \\
\frac{\alpha^2+\beta^2}{2} & \alpha & \beta &1-\frac{\alpha^2+\beta^2}{2}  \\
\end{array}\right).}
Its action on the polarization vectors is given by
\dis{W(\alpha,\beta)\epsilon_\pm^\mu=\epsilon_\pm^\mu+\frac{\alpha\mp i\beta}{\sqrt2}\frac{k^\mu}{\omega_k},\label{Eq:littleaction}}
which seems to be equivalent to the gauge transformation.
Indeed, this fact was used to introduce a gauge symmetry to the vector field even in the absence of charged particle \cite{Weinberg:1995mt}.
 Since the action of the non-compact little group on the vector field is a `gauge transformation', we expect that the residual symmetry emerging at null infinity generated by one of the two non-compact little group generators appears as the LGT charge. 
 The appearance of the same algebraic structure $ISO(2)$
 gives evidence to this assertion 
 that the gauge transformation generated by $\Pi_1$ appears as $\Delta Q_\varepsilon^{(s)}$ at null infinity. 
 In addition, comparing the action of  $\Delta Q_\varepsilon^{(s)}$ and $\Delta P_\varepsilon^{(s)}$ in eq. (\ref{Eq:JandQ}) and eq. (\ref{Eq:P}), we immediately find once again the common feature between the closed LGT algebra and the little group.
  Whereas $\Delta Q_\varepsilon^{(s)}$ acts on the left- and the right-handed helicity operators in the same way, i.e., helicity universal,  $\Delta P_\varepsilon^{(s)}$ acts differently on different helicity states.
  The action of the non-compact little group generators (eq. (\ref{Eq:littleaction})) also has the same property.
 The transformation generated by $\Pi_1$ is helicity universal but that generated by $\Pi_2$ is helicity distinguishing.
  
 Our discussion so far supports a close relation between the closed LGT algebra and  the little group for massless photons. 
 But still, the nature of $\Delta P_\varepsilon^{(s)}$ is not clear.
 Which type of gauge transformation distinguishes the helicities, and why it does not appear as a residual gauge symmetry at null infinity?

\subsection{Dual gauge transformation}

 In order to answer to the question raised in the previous section, we go back to the little group action on the polarization vector, eq. (\ref{Eq:littleaction}).
 Since both $\Pi_1$ and $\Pi_2$ generate the longitudinal polarization proportional to $k_\mu$,  we naively expect that these two 
 operators
 generate the gauge transformation.
 However, we find that the gauge transformation at null infinity is in fact helicity universal, and this is the reason why only $\Delta Q_\varepsilon^{(s)}$ is induced from the gauge transformation.
 To see this, let us consider the mode expansion of the gauge transformation parameter $\Lambda(x)$ in $A_\mu \to A_\mu+\partial_\mu \Lambda$.
 In terms of BRST cohomology, $\Lambda(x)$ in the gauge transformation corresponding to the ghost operator $c(x)$. 
 In the case of LGT, $\Lambda(x)$ is just a scalar function $\varepsilon(z, \bar{z})$.
 In any case, $\Lambda(x)$ has the free field expansion in the form of
 \dis{\Lambda(x) =\int\frac{d^3k}{(2\pi)^3(2\omega_k)}(c^\dagger(k)e^{-ik\cdot x}+c(k)e^{ik\cdot x}).}
 Using eq. (\ref{Eq:planewave}),  we find that 
   \dis{\partial_\mu & \Lambda(x)
   \\
   =& \int\frac{d\omega_k d^2z_k}{8\pi^2 r}
   \Big[-i(c^\dagger(k)e^{i\omega_k u}-c(k)e^{-i\omega_k u})
(\partial_\mu z_k\partial_{z_k}\delta^2(z-z_k)+\partial_\mu \bar{z}_k\partial_{{\bar z}_k}\delta^2(z-z_k)+\cdots)
\\
&\quad\quad +i(c^\dagger(k)e^{i\omega_k (u+2r)}-c(k)e^{-i\omega_k (u+2r)})
(\partial_\mu z_k\partial_{z_k}\delta^2(z+z_k)+\partial_\mu \bar{z}_k\partial_{{\bar z}_k}\delta^2(z+z_k)+\cdots)\Big],}
   thus
   \dis{\partial_z & \Lambda(x)
   \\
   &= \int\frac{d\omega_k d^2z_k}{8\pi^2 r}\Big[-i(c^\dagger(k)e^{i\omega_k u}-c(k)e^{-i\omega_k u})\partial_{z_k}\delta^2(z-z_k)
 \\
   &\quad\quad\quad\quad+i(c^\dagger(k)e^{i\omega_k (u+2r)}-c(k)e^{-i\omega_k (u+2r)})\partial_{z_k}\delta^2(z+z_k)\Big]
   \\
   &=\int\frac{d\omega_k }{8\pi^2 r}\Big[i(\partial_{z}c^\dagger(\vec{k})e^{i\omega_k u}-\partial_z c(\vec{k})e^{-i\omega_k u})
   -i(\partial_{z}c^\dagger(-\vec{k})e^{i\omega_k (u+2r)}-\partial_z c(-\vec{k})e^{-i\omega_k (u+2r)})\Big].}
   At null infinity, we take $r\to \infty$, and the $u$- and $r$-independent term consistent with the LGT comes from soft photon satisfying $\omega r \ll 1$ : the $\pm 2i\omega$ term in $e^{\pm i\omega(u+ 2r)}/r \simeq (1/r)(1\pm i\omega (u+2r) +\cdots)$.
   Hence, as a LGT, $\partial_z \Lambda$ appears to be a residual gauge transformation of $A_z$ with polarization $\epsilon_+=(\epsilon_-)^*$, while $\partial_{\bar z} \Lambda$ is a gauge transformation of $A_{\bar z}$ with polarization $\epsilon_-=(\epsilon_+)^*$, satisfying
\dis{& a_+ (\omega \hat{\mathbf{x}}_{z\bar{z}})
\to a_+(\omega \hat{\mathbf{x}}_{z\bar{z}})+\frac{2i \omega}{\sqrt{\gamma_{z \bar{z}}}}\partial_z c(-z, -\bar{z}),
\\
& a_- (\omega \hat{\mathbf{x}}_{z\bar{z}})
\to a_-(\omega \hat{\mathbf{x}}_{z\bar{z}})+\frac{2i \omega}{\sqrt{\gamma_{z \bar{z}}}}\partial_{\bar z} c(-z, -\bar{z}).\label{Eq:LGTtrans}}
By identifying $k_\mu=i\partial_\mu$, 
\footnote{in fact, as we go to $r\to \infty$, $\hat{k}\to \hat{x}$ and $k_z=\partial_z x^\mu k^\mu=0$.
But at least formally, we may regard $i\partial_{z/\bar{z}}$ as $k_{z/\bar{z}}$.}
we see that after absorbing the phase of the second term in eq. (\ref{Eq:LGTtrans}), the gauge transformation at null infinity is universal for both polarizations, which shares the same structure as $\Delta Q_\varepsilon^{(s)}$.

 Now, we recall that $\Delta Q_\varepsilon^{(s)}$ comes from the electric field $F_{u z/\bar{z}}$.
 When we describe the electromagnetic radiation, given an electric field in the circular polarization $\vec{E}=E_0(\vec{\epsilon}_+ + \vec{\epsilon}_-)e^{-i k\cdot x}$, the magnetic field is given by $\vec{B}=\hat{k}\times \vec{E}=E_0(i\vec{\epsilon}_--i\vec{\epsilon}_+)e^{-i k\cdot x}$ so that the two polarizations are distinguished by  the relative phase.
 From this, we notice that the helicitiy structure of $\Delta Q_\varepsilon^{(s)}$ and  $\Delta P_\varepsilon^{(s)}$ are those of the electric and the magnetic field, respectively.
 Indeed, the symmetry of the soft photon at null infinity 
 that is magnetic in nature 
 was already discussed in, e.g., \cite{Winicour:2014ska, Mao:2017axa, Strominger:2015bla}.
 Notably, \cite{Strominger:2015bla} suggests to consider such magnetic effect in terms of electromagnetic duality.
 This can be easily understood from the fact that under the electromagnetic duality, the roles of the electric field and the magnetic field are interchanged with each other.
 In our discussion, we have focused on the behavior of soft photons around null infinity, where any massive particle cannot reach. 
 Assuming the absence of massless charged particle, the Maxwell equations at null infinity are given by the sourceless form, $dF=d\widetilde{F}=0$, and electromagnetic duality becomes evident.
 In this case, just as gauge field and field strength are related by $F=dA$, we can introduce the `dual gauge field' $\widetilde{A}$ such that the dual electromagnetic field strength is written as $\widetilde{F}=d\widetilde{A}$.
 In addition, the `dual gauge transformation' under which $\widetilde{A}_\mu \to \widetilde{A}_\mu+\partial_\mu \Lambda$ can be imposed as a symmetry of QED at null infinity.
 Then it is natural to think of the emergent residual gauge symmetry at null infinity for the dual gauge transformation.
 Such `large dual gauge transformation'(LdGT) charge is, in analogous to LGT charge defined in eq. (\ref{Eq:LGTcharge}) \cite{Strominger:2015bla},
 \dis{P_\varepsilon  = -\int_S d^2z r^2 \gamma_{z\bar{z}}\varepsilon(z, \bar{z})\widetilde{F}_{ru}=-i\int_S d^2z \varepsilon(z, \bar{z}) F_{z\bar{z}}.}
 From this, the change of LdGT charge is given by
 \dis{\Delta P_\varepsilon &=-i\int_{-\infty}^{+\infty}du\int d^2z \varepsilon \partial_u (\partial_z A_{\bar z}-\partial_{\bar z}A_z)
 \\
 &=- \int d^2z \lim_{\omega\to 0}\frac{\omega \sqrt{\gamma_{z\bar{z}}}}{8\pi}\Big[i(\partial_z\varepsilon(z,\bar{z})a_+ (\omega\hat{x}) -\partial_{\bar z}\varepsilon(z, \bar{z})a_+^\dagger(\omega\hat{x}))
\\
&\quad\quad\quad -i(\partial_{\bar z}\varepsilon(z,\bar{z})a_-(\omega\hat{x})-\partial_z\varepsilon(z,\bar{z})a_-^\dagger(\omega\hat{x}))\Big],}
 the structure we expect from $\Delta P_\varepsilon^{(s)}$.
 Therefore, after taking the soft part of the LGT and the LdGT into account, the soft photon has three quantum numbers, i.e., its helicity, $\Delta Q_\varepsilon^{(s)}$, and $\Delta P_\varepsilon^{(s)}$. Together, these operators form a closed $ISO(2)$ algebra.
 
  This conclusion in turn motivates us to revisit the  meaning of the little group action, eq. (\ref{Eq:littleaction}).
  As mentioned previously, the appearance of the longitudinal mode as a result of the non-compact little group action has been regarded as a way to introduce a gauge symmetry to the system with gauge bosons only: having charged matter is not an essential ingredient for the gauge symmetry.
  However, so far as the vector field is concerned, the system has not only a gauge symmetry, but also a dual gauge symmetry.
  An interchange between the electric and the magnetic fields under the dual gauge transformation is reflected in the action of $\Pi_1$ and $\Pi_2$, such that while the action of $\Pi_1$ is helicity universal (electric in nature), the action of $\Pi_2$ distinguishes the helicity (magnetic in nature).
  Looking at how the gauge and the dual gauge transformation appears at null infinity, their residual symmetry appear as LGT and LdGT.

\section{Detecting the LGT charge via the Electric Aharonov-Bohm Effect}
\label{Sec:AB}
\setcounter{equation}{0}                                                                                                                                                                                                                                                                                                                                                                                                                                                                                                                                                                                                                                      

 One interesting aspect of LGT is its connection to the memory effect.
  Originally, the memory effect was defined in the context of gravitational waves, as a permanent change between two gravitational wave detectors after the gravitational wave passes through.
  Analogously, we can think of the electromagnetic memory as describing a permanent change of the {\it motion of charged particle} after the passage of an electromagnetic wave \cite{Bieri:2013hqa}.
   Another point of view in interpreting the memory effect is the change encoded in the {\it gauge field} resulting from the passage of the wave.
This later viewpoint turns out to be useful in relating the memory effect with the soft theorem.
   In the case of the gravitational memory, the change of curvature is restricted to be a subleading effect, thus maintaining the asymptotic flatness.
 That means that whereas two geometries before and after the passage of the gravitational wave  are distinct, the difference in curvature is very small, and hence the same flat spacetime background is dominant around null infinity.
 This is exactly the setting where the super-translation/rotation in the BMS group emerges as the approximate symmetry at null infinity: different asymptotically flat geometries are connected by the BMS transformation.
 Moreover, asymptotic flatness is maintained if only the soft gravitons affect the dynamics around null infinity. 
 Hence, asymptotic symmetry and the memory effect are closely relevant to the physics of soft gravitons.
 
  We can apply the discussion above to electrodynamics.
  Basically, the memory effect is characterized by the integration of the electric field over an infinite range of time.
  As discussed in  \cite{Bieri:2013hqa}, this integral can be expressed as a jump of the vector field $A_{\mu}$ before and after the passage of the electromagnetic wave.
   To have a non-vanishing memory effect, 
   the vector field $A_{\mu}$ before and after need to be gauge inequivalent, but such inequivalence is a subleading effect at null infinity, where the leading order vector field is given by a pure gauge.  
    Then we can say that two vector fields are approximate gauge equivalent, connected by the LGT.
    Soft photons are important in this regard, as they just alter the subleading effects of the vector field, which are suppressed by some positive power of $r$. 

    Two gauge equivalent fields can be distinguished by the Aharonov-Bohm effect. 
    Hence, we can utilize the Aharonov-Bohm effect to distinguish two gauge fields connected by a LGT, where the inequivalence is suppressed by ${\cal O}(r^{-n})$ ($n>0$).
    The suggestion of \cite{Susskind:2015hpa} is based on such considerations.
   The Aharonov-Bohm effect in \cite{Susskind:2015hpa} is the conventional one, trading the phase difference of test charged particles moving  along different trajectories
   with the magnetic flux.
  On the other hand, our discussion so far shows that the LGT   has the nature of an electric field, rather than a magnetic field.
  This will be evident shortly as we show that the combination of the gauge fields contributing to the LGT charge $\Delta Q_\varepsilon^{(s)}$ is equivalent to the scalar potential
  \footnote{  Note that, in \cite{Susskind:2015hpa}, the temporal gauge where the scalar potential is zero was taken.}.
  That means that a more direct measure of the electromagnetic memory effect is the potential difference made by the passage of  soft photons, which can be measured by the electric Aharonov-Bohm effect.

 To see the discussion above in detail, let us go back to the Maxwell equation, eq. (\ref{Eq:Maxwell}).
 In the usual $(t, r', z, \bar{z})$ coordinate, the scalar and the vector potential are given by $A^\mu=(A^t, \vec{A})=(\phi, \vec{A})$. 
In the retarded time coordinate where $t=u+r$ ad $r'=r$, the vector potentials are related by $A_u=A_t=-\phi$ and $A_r=A_t+A_{r'}=-\phi+A_{r'}$. 
Now, consider the radial gauge $A_r=0$.
Assuming no massless charged particle at null infinity, $J^r=0$.
Then,
\dis{\partial_u F_{ru}=\partial_u \partial_r A_u=-\partial_u\partial_r\phi =-\frac{1}{r^2\gamma_{z\bar{z}}}\partial_u(\partial_zA_{\bar z}+\partial_{\bar z}A_z),}
or
\dis{\partial_u \phi=-\frac{1}{r \gamma_{z\bar{z}} }  \partial_u (\partial_zA_{\bar z}+\partial_{\bar z}A_z).\label{Eq:potmem}}
Therefore, the difference between the scalar potential $\phi$ at $u=+\infty$ and at $-\infty$ corresponds to
\dis{\phi(u=+\infty)-\phi(u=-\infty)=-\frac{1}{r\gamma_{z\bar{z}}}\int^{+\infty}_{-\infty}du  \partial_u (\partial_zA_{\bar z}+\partial_{\bar z}A_z).\label{Eq:potential}}
The RHS is equivalent (after scaling out the factor of $1/r$) 
to the LGT charge $\Delta Q_\varepsilon^{(s)}$ with $\varepsilon=\delta^2(z_0-z)/\gamma_{z\bar{z}}$ 
at a specific angular parameter $z_0$.

  In the Lorenz gauge $\partial_\mu A^\mu=0$, since
 \dis{\partial_u A_r=\frac{1}{r^2}\partial_r[r^2(-A_u+A_r)]+\frac{1}{r^2\gamma_{z\bar{z}}}[\partial_z A_{\bar z}+\partial_{\bar z}A_z],}
 the Maxwell equations become
 \dis{\partial_u F_{ru}&=\partial_u \partial_r A_u-\partial_u^2 A_r 
 \\
 &= \partial_u \partial_r A_u-\partial_u \Big[\frac{1}{r^2}\partial_r[r^2(-A_u+A_r)]+\frac{1}{r^2\gamma_{z\bar{z}}}[\partial_z A_{\bar z}+\partial_{\bar z}A_z] \Big]
=-\frac{1}{r^2\gamma_{z\bar{z}}}\partial_u(\partial_zA_{\bar z}+\partial_{\bar z}A_z).}
 Comparing the last two equalities, we find that the electric field $F_{uz/\bar{z}}$'s contribution to the memory effect vanishes.
 Instead, the gauge fixing condition itself relates the scalar potential to the memory effect.
 Following the boundary condition eq. (\ref{Eq:boundary}), we neglect $A_r$ compared to $A_u$, then the Lorenz gauge fixing condition becomes
 \dis{0&=\partial_\mu A^\mu \simeq -\partial_r A_u+\frac{1}{r^2\gamma_{z\bar{z}}}(\partial_{\bar z} A_z+\partial_z A_{\bar{z}} )
=\partial_r \phi+\frac{1}{r^2\gamma_{z\bar{z}}}(\partial_{\bar z} A_z+\partial_z A_{\bar{z}} ),}
 to give eq. (\ref{Eq:potmem}) again with the sign flipped.
 
  In any case, the potential difference between the long time separation comes from the soft photon reaching null infinity, or a detector far away from the source, so it is a direct measure of the LGT charge with the parameter $\varepsilon$ localized at a specific angle.
  If we start with the universal initial potential i.e., $\phi(u=-\infty)$=const. everywhere on the spherical surface, after a long enough time, $\phi(u=\infty, z, \bar{z})$ becomes different for each angular coordinates parametrized by $z$ as different number of soft photons reach the spherical surface.  
   
   \begin{figure}[t]
 \begin{center}
   \includegraphics[width=10cm]{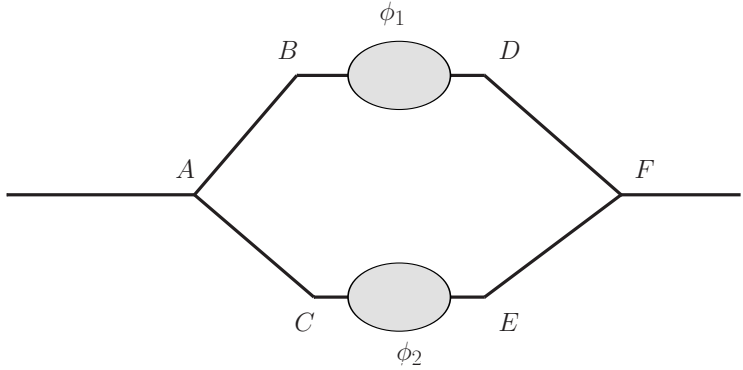}
  \vspace{-1em}
 \end{center}
 \caption{Experimental setup for the electric Aharonov-Bohm effect, as shown in \cite{Aharonov:1959fk}. }
 \label{Fig:electricAB}
\end{figure}

   Even locally, we may measure $\phi(u=\infty, z, \bar{z})-\phi(u=-\infty, z, \bar{z})$ at some specific $z$ in the following way:
First, we connect the external voltage generator to both the BD region and the CE region in Fig. \ref{Fig:electricAB} to make the initial potentials in the two regions take the same value, $\phi_2=\phi(u=-\infty)$.
 Next, while keeping the connection between the voltage generator and  the CE region, we connect the BD region to the ground such that the potential in the BD region is changed to $\phi_1=\phi(u=+\infty)$ resulting from the passage of soft photons.
 Then by sending test charges, we can compare the potentials in the BD region and the CE region to measure the soft part of the LGT charge $\Delta Q_\varepsilon^{(s)}$.
 This is how the so-called electric Aharonov-Bohm effect works. 
  
  Of course, such experimental setup is an ideal one. 
In real situations, there are several corrections arising from
e.g., having a finite $r$, a finite time interval, and a limitation of the resolution of the angular coordinate $z$.
The finite $r$ correction is controllable as every argument so far was made based on an ${\cal O}(1/r)$ expansion. Hence, we comment instead on the effects of a finite time interval and a limited angular resolution.
Both 
effects
have to do with the sharpness of the delta function, so we 
can focus our discussion on the former.
For this purpose, suppose instead of $\infty$, we use a finite but large boundary value $U$:
  \dis{\phi(u=+U, z_0)-\phi(u=-U, z_0)=-\frac{1}{r\gamma_{z\bar{z}}(z_0)}\int^{+U}_{-U}du  \partial_u (\partial_{z}A_{\bar{z}}(u, z_0)+\partial_{\bar{z}}A_{z}(u, z_0)).\label{Eq:potentialdiff}}
  But we find that the potential difference is insensitive to $U$ when $U$ is very large.
  To see this, consider the $\omega$ integration using regulators $U$ and $\epsilon$ as 
   \dis{\int_{-U}^{+U} du e^{i\omega u}&=\lim_{\epsilon \to 0}\Big[\int_{0}^{+U}du e^{iu(\omega+i\epsilon)}+\int_{-U}^{0}du e^{iu(\omega-i\epsilon)}
  \Big]
  \\
  &=\lim_{\epsilon \to 0} e^{-\epsilon U}\Big[\frac{e^{i\omega U}}{i \omega-\epsilon}-\frac{e^{-i\omega U}}{i\omega+\epsilon}\Big]
  +\Big[-\frac{1}{i\omega-\epsilon}+\frac{1}{i\omega+\epsilon}\Big]
  \\
  &=\lim_{\epsilon \to 0}  e^{-\epsilon U}\Big[\frac{2\omega \sin(\omega U)}{\omega^2+\epsilon^2}-\frac{2\epsilon \cos(\omega U)}{\omega^2+\epsilon^2}\Big]+\frac{2\epsilon}{\omega^2+\epsilon^2}
.}
The first term with the bracket (i.e., the oscillating term)  rapidly attenuates as $\epsilon U\gg 1$.
The last term gives the Dirac delta function $2\pi\delta(\omega)$ through
\dis{\delta(x)=\frac{1}{\pi}\lim_{\epsilon\to 0}\frac{\epsilon}{\omega^2+\epsilon^2}.}
Here, $\epsilon$ is interpreted as the resolution of the frequency.
Hence, so far as the measurement time scale $U$ is much longer than $1/\epsilon$, the oscillating term can be neglected.
Since the $U$ dependence is contained in the oscillating term only, we may neglect the $U$ effects.
From now on, we change our notation: in order to emphasize the role of $\epsilon$ as a frequency resolution, we denote $\epsilon \equiv \Delta \omega$. 
  
  Using the mode expansion eq. (\ref{Eq:gaugemode}), the RHS of eq. (\ref{Eq:potentialdiff}) becomes 
  \dis{&-\frac{1}{r\gamma_{z\bar{z}}(z_0)}\int^{+U}_{-U}du  \partial_u (\partial_{z}A_{\bar{z}}(u, z_0)+\partial_{\bar{z}}A_{z}(u, z_0))
\\
  &=
  -\int d^2 z\frac{\delta^2(z_0-z)}{r\gamma_{z\bar{z}}(z)}\int^{+U}_{-U}du  \partial_u (\partial_{z}A_{\bar{z}}(u, z)+\partial_{\bar{z}}A_{z}(u, z))
\\
&=-\int d^2 z\sqrt{ \gamma_{z\bar{z}}}  \int_0^{\Lambda} \frac{d\omega}{8\pi^2 r}\omega \frac{2\Delta\omega}{\omega^2+\Delta\omega^2} \Big[\partial_{z}\Big(\frac{\delta^2(z_0-z)}{\gamma_{z\bar{z}}}\Big)(a_+(\omega\hat{x})+a_-^\dagger(\omega\hat{x}))+ 
\\
&\quad\quad\quad\quad\quad + \partial_{\bar{z}}\Big(\frac{\delta^2(z_0-z)}{\gamma_{z\bar{z}}}\Big)(a_-(\omega\hat{x})+a_+^\dagger(\omega\hat{x}))\Big]
\\
&= -\int d^2 z\sqrt{ \gamma_{z\bar{z}}}\frac{1}{8\pi^2 r}\Delta\omega\log\Big(1+\frac{\Lambda^2}{\Delta\omega^2}\Big)\Big[\partial_{z}\Big(\frac{\delta^2(z_0-z)}{\gamma_{z\bar{z}}}\Big)(a_+(\omega\hat{x})+a_-^\dagger(\omega\hat{x}))+ 
\\
&\quad\quad\quad\quad\quad + \partial_{\bar{z}}\Big(\frac{\delta^2(z_0-z)}{\gamma_{z\bar{z}}}\Big)(a_-(\omega\hat{x})+a_+^\dagger(\omega\hat{x}))\Big].}
  As already implied in eq. (\ref{Eq:LGTharge}), the potential difference vanishes for the exact zero frequency, $\Delta \omega=0$. 
 This is an expected result since zero energy transferred by photons does not change the potential energy at all.
 The change in potential arises from extremely small but not exactly zero frequency photons, which we conventionally refer to `soft' photons.
 Of course, as $U\to\infty$, high frequency effect would be strongly suppressed.

 We note here that in detecting both the electric and the magnetic Aharonov-Bohm effects, we need to use a test charge.
As soon as we introduce a test charge, the electromagnetic duality, which arises in the sourceless Maxwell equation, is spoiled.
This breaks the dual gauge symmetry explicitly, and as a result the LdGT generated by $\Delta P_\varepsilon$, which has a direct connection to the well-known magnetic Aharonov-Bohm effect, is no longer the symmetry.
 Since the magnetic flux can be measured with only a non-negligible amount of charge of the test particle, the measurement of $\Delta P_\varepsilon$ 
 becomes unreliable.


\section{Conclusions}
\label{Sec:conclusion}

 In this article, we have emphasized the electric field nature of  the large gauge transformation charge. 
 If we neglect the charged matter, electromagnetic duality emerges, and large dual gauge transformation, governed by the magnetic field nature needs to be taken into account.
 Together with the helicity operator, these two charges form an $ISO(2)$ algebra, which has the same algebraic structure as the little group of massless particles.
 This is consistent with the little group action on the polarization vector of the vector field.
 In this regard, we can make a more precise interpretation of the little group action on the polarization vector: while the helicity universal part is the gauge symmetry of the pure Abelian gauge theory, the helicity distinguishing part is then the dual gauge transformation.
 Moreover, we find that the electric field nature of the large gauge transformation
 suggests an interesting possibility that the large gauge transformation can be measured by the electric Aharonov-Bohm effect. 
  
 While we focus our investigation on QED, we expect 
 a similar
 structure to appear in the
 gravitational case.
Indeed, the gravitational memory effect relevant to supertranslation is encoded in the electric field part of the Weyl tensor, as emphasized in, e.g., \cite{Madler:2016ggp}. Moreover, the electric part of the Weyl tensor comprises of the Bondi mass aspect whose angular integration provides the BMS charge for supertranslation \cite{Flanagan:2015pxa}.
As the BMS charges also create/annihilate soft gravitons at null infinity, they do not commute with the helicity operator, and another soft photon generator made up of the magnetic part of the Weyl tensor may be included to close the algebra. 
 On the other hand, it was recently suggested that asymptotic symmetries in gravity are associated with the Berry phase \cite{Oblak:2017ect, Oblak:2017ptc}, from which we may design detection of the gravitational memory effect through the electric Aharonov-Bohm-like effect. 
 We hope to return to these interesting issues about the gravitational memory effect in the future.

\subsection*{Acknowledgements}

We thank Daniel Chung, Deog-Ki Hong, Toshifumi Noumi
for discussions.
MS is grateful to the String Theory and Theoretical Cosmology research group, Department of Physics at the University of Wisconsin-Madison for hospitality during his visit.
The work of MS is supported by the National Research Foundation of Korea (NRF-2016R1C1B2015225). 
The work of YH is supported in part by the Grant-in-Aid for JSPS Fellows No.16J06151.
GS is supported in part by the DOE grant DE-SC0017647 and the Kellett Award of the University of Wisconsin.




\end{document}